\documentclass[11pt]{article}

\usepackage{amsmath}
\usepackage{amssymb}
\usepackage{graphicx}
\usepackage{subfigure}

\def\middlespace {\smallskipamount=5.625pt plus1.5pt minus1.5pt
                  \medskipamount=11.25pt plus3pt minus3pt
                  \bigskipamount=22.5pt plus6pt minus6pt
                  \normalbaselineskip=22.5pt plus0pt minus0pt
                  \normallineskip=1pt
                  \normallineskiplimit=0pt
                  \jot=5.625pt
                  {\def\smallskip {\vskip\smallskipamount}}
                  {\def\medskip   {\vskip\medskipamount}}
                  {\def\bigskip   {\vskip\bigskipamount}}
                  {\setbox\strutbox=\hbox{\vrule
                    height15.75pt depth6.75pt width 0pt}}
                  \parskip 11.25pt
                  \normalbaselines}

\begin{document}
\ \vskip 1.0 in

\begin{center}
 { \Large {\bf Quantum Theory, Noncommutative Gravity,}}

\smallskip

{\Large {\bf  and the Cosmological Constant Problem}}

\vskip 0.2 in

\smallskip

\bigskip

\bigskip

\bigskip

{{\large
{\bf T. P. Singh\footnote{e-mail address: tpsingh@tifr.res.in} 
} 
}}

\medskip

{\it Tata Institute of Fundamental Research,}\\
{\it Homi Bhabha Road, Mumbai 400 005, India.}\\
\medskip
\vskip 0.5cm
\end{center}

\vskip 1.0 in

\begin{abstract}

\noindent The cosmological constant problem is principally concerned with trying to understand how the
zero-point energy of quantum fields contributes to gravity. Here we take the approach that by addressing
a fundamental unresolved issue in quantum theory we can gain a better understanding of the problem.
Our starting point is the observation that the notion of classical time is external to quantum mechanics. Hence there must exist an
equivalent reformulation of quantum mechanics which does not refer to an external classical time.
Such a reformulation is a limiting case of a more general quantum theory which becomes
nonlinear on the Planck mass/energy scale. The nonlinearity gives rise to a quantum-classical duality which maps  a `strongly quantum, weakly gravitational' dynamics to a `weakly quantum, strongly gravitational' dynamics. This duality predicts the existence of a tiny nonzero cosmological constant of the order of the square of the Hubble constant, which could be a possible source for the observed cosmic acceleration. Such a nonlinearity could also be responsible for the collapse of the wave-function during a quantum measurement. We conclude by suggesting that the idea for the origin of dark energy proposed in this paper can be tested indirectly in the laboratory by a detailed examination
of the process of quantum measurement.

\vskip 1.0 in

\end{abstract}

\newpage

\section{Introduction}

The observed notion of time, with which we are so familiar, is external to quantum mechanics. It is part of
a classical spacetime geometry, which comprises of a spacetime manifold and the metric. The metric is determined
by classical matter fields via the field equations of general relativity. In principle, the Universe could be
in a state in which there are no classical matter fields, but only quantum fields. In such a situation, the
metric of the Universe will in general no longer be classical, but will undergo quantum fluctuations. It is
known from the Einstein hole argument that in order for the spacetime manifold to have a physically meaningful
point structure, a well-determined classical metric (which is a solution of the Einstein equations) must reside on the manifold. When the metric is undergoing quantum fluctuations, the point structure of the spacetime manifold
is destroyed, and one no longer has a classical notion of time \cite{singh}.

Nonetheless, one should be able to describe the dynamics of a quantum system, even if an external classical time
is not available. Such a description should become equivalent to standard quantum mechanics as and when a dominant
part of the Universe becomes classical, so that a classical time now exists. In arguing for the existence of
such a reformulation one is led to conclude that standard linear quantum theory is a limiting case of a more
general quantum theory which is nonlinear on the Planck mass/energy scale \cite{singh2}. This conclusion is 
{\it independent} of any specific mathematical structure which one would like to use to develop the
reformulation.

What is our most reliable guideline towards the construction of such a reformulation of quantum mechanics?
A natural mathematical structure which forgoes the point structure of spacetime is a noncommutative spacetime.
We construct the reformulation by pursuing the following proposal : in the reformulation, relativistic quantum
mechanics is the same theory as noncommutative special relativity, with a specific set of commutation relations
imposed on noncommuting coordinates and momenta. The physical principle is that the basic laws are invariant under
`inertial' coordinate transformations of noncommuting coordinates. One is naturally led to attach an antisymmetric part to the Minkowski metric. The theory is supposed to describe dynamics when gravity can be neglected (like in special relativity). In the present context, this amounts to the requirement that the total mass energy in the
system be much smaller than Planck mass/Planck energy. As and when an external time becomes available, this
reformulation should become equivalent to standard quantum mechanics. These aspects will be described in
Section 2.

The nonlinear generalization of this reformulation, a hitherto unnoticed feature which arises naturally, describes
the dynamics of the system when its energy becomes comparable to Planck energy. The Schr\"{o}dinger equation becomes nonlinear and the gravitational dynamics is now a noncommutative general relativity. The physical
principle now is that basic laws are invariant under general coordinate transformations of noncommuting coordinates.
This is supposed to generalize general covariance to the noncommutative case. When the mass-energy becomes much
larger than the Planck scale, the dynamics is assumed to reduce to classical general relativity, and classical
mechanics. This is discussed in Section 3.

The presence of the nonlinearity has two important consequences. Firstly, the antisymmetric part of the
gravitational field associated with this nonlinearity suggests the existence of a quantum-classical duality,
as a consequence of which one can match a dominantly quantum sector of the theory to a dominantly classical sector.
This is the subject matter of Section 4. In turn this helps us understand why the cosmological constant should be non-zero and yet have the very small value it does. This is the main part of the paper, and will be presented in Section 5.

The second important consequence of the nonlinearity has to do with the nonlinearity in the Schr\"{o}dinger
equation, which becomes relevant in the vicinity of the Planck mass scale. This can lead to a breakdown of
quantum superposition, and could lead to the collapse of the wave-function during a quantum measurement.
What is important for us here is that the parameters influencing the collapse of the wave-function are in principle measurable in the laboratory. These are the same parameters which are responsible for the existence of the
quantum classical duality, and for the non-zero value of the cosmological constant. Thus our explanation
for the origin of the dark energy is in principle testable experimentally, via the quantum measurement process.
This aspect is investigated in Section 6.  
 
The arguments of this paper suggest that a dynamically evolving `cosmological-constant like' term is present throughout the history of the Universe. At any given epoch such a term is supposedly of the order of the square of the Hubble constant at that epoch. The cosmological viability of such a scenario will be discussed in Section 7. 

In this paper we have attempted to keep the discussion compact, so as to provide an essential overview of the arguments. More detailed discussions can be found in \cite{singh}, \cite{singh2} and \cite{singh3}.

\section{Quantum mechanics as a noncommutative special relativity}
The quantum dynamics of a relativistic particle of mass $m\ll m_{Pl}$
is described here as a noncommutative special relativity. Gravity is neglected in this small mass limit since
this approximation is equivalent to setting $G\rightarrow 0$. 
We outline here a proposal for the desired reformulation, using the illustrative case of a two-dimensional noncommutative spacetime described by coordinates $(\hat{x},\hat{t})$. It should be said at the outset that our treatment is heuristic, and a rigorous mathematical description remains to be developed. We assume that associated with the 2-d noncommutative spacetime there is a line element
\begin{equation}
d\hat{s}^{2}=\hat{\eta}_{\mu\nu}d\hat{x}^{\mu}d\hat{x}^{\nu}\equiv
d\hat{t}^{2}-d\hat{x}^{2}
+d\hat{t}d\hat{x}-d\hat{x}d\hat{t},
\label{lin}
\end{equation}
which has an antisymmetric component. We call such a spacetime a quantum Minkowski spacetime, and the noncommuting coordinates $\hat{t}, \hat{x}$ are assumed to obey the commutation relations 
\begin{equation}
[\hat{t},\hat{x}]=f^{-1}(\hat{p}^{t},\hat{p}^{x}), \qquad [\hat{p}^{t}, \hat{p}^{x}]=f(\hat{p}^{t},\hat{p}^{x}).
\label{commu}
\end{equation}
We will comment on the function $f$ shortly.

We assume that a suitable differential calculus can be defined on this 
spacetime. Then, in analogy with special relativity, we introduce a velocity
$\hat{u}^{i}=d\hat{x}^{i}/d\hat{s}$ and a momentum $\hat{p}^{i}=m\hat{u}^{i}$.
It is evident from the form of the line-element (\ref{lin}) that the following Casimir relation holds
\begin{equation}
(\hat{p}^{t})^{2}-(\hat{p}^{x})^{2} + 
\hat{p}^{t}\hat{p}^{x} - \hat{p}^{x}\hat{p}^{t}  = m^{2}.
\label{nce}
\end{equation} 

The specific structure of the commutation relations above is such that the momenta, as well as the coordinates,
do not commute with each other. Moreover, while $f$ appears in one of the relations, it is $f^{-1}$ which
appears in the other relation. This is motivated by the expectation that one should be able to derive
the uncertainty relations of quantum theory, and the quantum commutation relation $[q,p]=i\hbar$ from these
underlying relations \cite{singh2}. 

The function $f$ in (\ref{commu}) has to be so chosen that the momenta commute with the Casimir relation.  
It is easy to show that in fact there is no non-trivial solution in two dimensions;
the  only solution is $f=0$, which is clearly not of interest. However, in dimensions three or higher
there appears to be no constraint that $f=0$, although the exact form of $f$ remains to be found.
Our subsequent discussion here does not depend on the form of $f$, and it suffices to use the 2-d example
to illustrate our ideas. 

Dynamics is defined by assuming that the momenta are gradients of a 
{\it complex} action $\hat{S}$. This converts the Casimir relation into a noncommutative Hamilton-Jacobi equation, which is the equation of motion. This is the theory we call a noncommutative special relativity.

As and when an external classical spacetime $(x,t)$ becomes available, the Klein-Gordon equation of standard linear quantum mechanics can be recovered from this reformulation by the correspondence rule
\begin{equation}
(\hat{p}^{t})^{2}-(\hat{p}^{x})^{2} + 
\hat{p}^{t}\hat{p}^{x} - \hat{p}^{x}\hat{p}^{t}  = ({p}^{t})^{2}-({p}^{x})^{2}
 + i\hbar {\partial p^{\mu}\over \partial x^{\mu}}.
\label{nceq}
\end{equation}
The justification for this rule has been discussed in \cite{singh2}. On the right hand side of this equation, the momenta are again defined as the gradients of a complex action $S$, and the wave-function defined as 
$\psi\equiv e^{iS/\hbar}$. Substituting for the wave function on the right hand side of (\ref{nceq}) and equating this expression to $m^2$ leads to the Klein-Gordon equation. In this sense one can recover standard quantum mechanics 
 from an underlying formulation as a noncommutative special relativity.

\section{A noncommutative general relativity}
When the mass of the particle becomes comparable to Planck mass, its self-gravity can no longer be neglected.The noncommutative line-element (\ref{lin}) is
modified to the curved noncommutative line-element
\begin{equation}
ds^{2}=\hat{h}_{\mu\nu}d\hat{x}^{\mu}d\hat{x}^{\nu}\equiv
\hat{g}_{tt}d\hat{t}^{2}-\hat{g}_{xx}d\hat{x}^{2}
+\hat\theta[d\hat{t}d\hat{x}-d\hat{x}d\hat{t}].
\label{linc2}
\end{equation}
Correspondingly, the Casimir relation (\ref{nce}) is generalized to 
\begin{equation}
\label{nceq2}\hat g_{tt}(\hat p^t)^2-\hat g_{xx}(\hat p^x)^2+\hat \theta
\left( \hat p^t\hat p^x-\hat p^x\hat p^t\right) =m^2
\end{equation}
and the correspondence rule (\ref{nceq})to
\begin{equation}
\label{corr}\hat g_{tt}(\hat p^t)^2-\hat g_{xx}(\hat p^x)^2+\hat \theta
\left( \hat p^t\hat p^x-\hat p^x\hat p^t\right)=g_{tt}({p}^t)^2-g_{xx}({p}%
^x)^2+i\hbar \theta {\frac{\partial p^\mu }{\partial x^\mu }}.
\end{equation}
It is important now to note that if one rewrites this Hamilton-Jacobi equation in terms of the wave-function, one no longer gets the linear Klein-Gordon equation. This is because the metric appears in the equation. In the simplest case, where $\theta$ is a function of $m/m_{Pl}$, and the diagonal components of the
metric are approximated to unity, we get the equation of motion
\begin{equation}
\label{hjcn}\left( {\frac{\partial {S}}{\partial t}}\right) ^2-\left( {\frac{%
\partial {S}}{\partial x}}\right) ^2-i\hbar \theta (m/m_{Pl}) \left( {\frac{\partial ^2S%
}{\partial t^2}}-{\frac{\partial ^2S}{\partial x^2}}\right) =m^2
\end{equation}
which is equivalent to a nonlinear Klein-Gordon equation \cite{singh2}.

The noncommutative metric is assumed to
obey a noncommutative generalization of Einstein equations, with the property that $\theta(m/m_{Pl})$
goes to one for $m\ll m_{Pl}$, and to zero for $m\gg m_{Pl}$. Also, as $\theta(m/m_{Pl})\rightarrow 0$
one recovers classical mechanics, and in the limit $\theta\rightarrow 1$ standard linear quantum mechanics is recovered. 

In the mesoscopic domain, where $\theta$ is away from these limits and the mass $m$ is comparable to Planck mass, both quantum and gravitational features can be defined simultaneously, and new physics arises. The antisymmetric component $\theta$ of the gravitational field plays a crucial role in what follows.  

\section{A proposed quantum-classical duality}

\noindent{\bf Motivation for the duality:} In general relativity, the Schwarzschild radius $R_S=2Gm/c^2$ of a particle of mass $m$ can be written in Planck units as $R_{SP}\equiv R_S/L_{Pl}=2m/m_{Pl}$, where $L_{Pl}$ is Planck length and $m_{Pl} \sim 10^{-5}$ gm is the Planck mass.
If the same particle were to be treated, not according to general relativity, but according to relativistic quantum mechanics, then one-half of the Compton wavelength $R_C=h/mc$ of the particle can be written in Planck units as $R_{CP}\equiv R_C/2L_{Pl}= m_{Pl}/2m$.  
The fact that the product $R_{SP}R_{CP}=1$ is a universal constant cannot be a coincidence; however it cannot be explained in the existing theoretical framework of general relativity (because herein $h=0$) and quantum mechanics (because herein $G=0$). 

One could attempt to trivialize this observation by saying that in general relativity the only length scale
that can be constructed {\it is} proportional to mass and in relativistic quantum theory the only length scale that
can be constructed {\it is} inversely proportional to mass. However, what is non-trivial is that both these
length scales have a fundamental physical meaning attached to them. Hence their inverse relation to each other
does call for an explanation, and is a signal that both general relativity and relativistic quantum theory
must be limiting cases of a deeper underlying theory. In fact we have argued for the existence of such a theory
in the previous section for entirely different reasons.
  
The only plausible way to explain this inverse relation is to propose a duality between a pair of solutions
of the theory - a duality which maps the Schwarzschild radius for the first solution to the Compton wavelength 
for the second solution. Hence we propose and justify the following quantum-classical duality:

\smallskip

{\it
\noindent The weakly quantum, strongly gravitational dynamics of a particle of mass $m_c\gg m_{Pl}$ is dual to the strongly quantum, weakly gravitational dynamics of a particle of mass $m_q=m_{Pl}^2/m_c\ll m_{pl}$.}

\smallskip

It follows that the dimensionless Schwarzschild radius $R_{SP}$ of $m_c$ is four times the dimensionless
Compton-wavelength $R_{CP}$ of $m_q$. 

The origin of this duality lies in the requirement that there be a reformulation of quantum mechanics
which does not refer to an external classical spacetime manifold. The implied nonlinearity leads to a quantum
gravity theory of which general relativity and quantum theory are natural approximations, and the duality
is inevitable. Its existence does not depend on the use of nonommutative geometry for the mathematical
formulation of the theory. The use of noncommutativity serves to illustrate and justify the duality.

The Planck mass demarcates the dominantly quantum domain $m < m_{Pl}$ from the dominantly classical
domain $m > m_{Pl}$ and is responsible for the quantum-classical duality. 
As is evident from (\ref{hjcn}), the effective Planck's constant is $\hbar\theta(m/m_{Pl})$, going to zero for large masses, and to $\hbar$ for small masses, as expected. Similarly,  the effective Newton's gravitational constant is expected to be  $G(1-\theta(m/m_{Pl}))$, going to zero for small masses, and to $G$ for large masses. 

 Thus the parameter space $\theta\approx 1$ is strongly quantum and weakly gravitational, whereas
$\theta\approx 0$ is weakly quantum and strongly gravitational. The Compton wavelength $R_{CP}$ for
a particle of mass $m_q$ gets modified to $R_{CE}\equiv R_{CP}\theta(m_q/m_{Pl})$ and the Schwarzschild radius $R_{SP}$ for a mass $m_c$ gets modified to $R_{SE}\equiv R_{SP}(1-\theta(m_c/m_{Pl}))$. 
We propose
that the dynamics of a mass $m_q\ll m_{Pl}$ is dual to the dynamics of a mass 
$m_c\gg m_{Pl}$ if $R_{SE}(m_c)=4R_{CE}(m_q)$. This holds if $m_c=m_{Pl}^2/m_q$ and 
\begin{equation} 
\theta(m/m_{Pl}) + \theta(m_{Pl}/m) = 1. 
\label{thetacon}
\end{equation}
If (\ref{thetacon}) holds, the solution for the dynamics for a particle of mass $m_c$ can be obtained
by first finding the solutions of (\ref{hjcn}) for mass $m_q$, and then replacing $\theta(m_q/m_{Pl})$ by $1-\theta(m_{Pl}/m_{q})$, and  finally writing $m_c$ instead of $m_q$, wherever $m_q$ appears. 

We can deduce the functional form of $\theta(m/m_{Pl})$ by noting that the contribution of the symmetric part of the metric, $g_{ik}$,
to the curvature, grows as $m$, whereas the contribution of the antisymmetric part $\theta$ must fall with growing $m$. This suggests that $1/\theta$ grows linearly with $m$; thus
\begin{equation}
\frac{1}{\theta(m/m_{Pl})}= a(m/m_{Pl})+b,
\end{equation} 
and $\theta(0)=1$ implies $b=1$; and  we set $a=1$ since this simply defines $m_{Pl}$ as the scaling mass. Hence we get $\theta(m/m_{Pl})=1/(1+m/m_{Pl})$, which satisfies (\ref{thetacon}) and thus establishes the duality.
The mapping $m\rightarrow 1/m$ interchanges the two fundamental length scales in the solution : Compton wavelength
and Schwarzschild radius.

The duality we
observe is holographic, by virtue of the above-mentioned relation $R_{SE}(m_c)=4R_{CE}(m_q)$. Thus, the number of degrees of freedom $N$ that a quantum field associated with the particle $m_q$ possesses (bulk property) should be of the order of the area of the horizon of the dual black hole in Planck units (boundary property), i.e. $N\sim m_{Pl}^2/m_{q}^2$. This value of $N$ could be interpreted as follows : the infinite number of degrees of freedom
associated with a quantum field in the flat spacetime continuum limit (when no artificial high-energy cut-off
has been imposed) has been replaced by this finite value. More correctly however, the effective number of degrees of
freedom is actually of the order $m_{Pl}/m_{q}$, because we have $m_{q}\ll m_{Pl}$ and so the highest energy
associated with a mode of the quantum field cannot be more than Planck mass.

In summary, we see here a new picture for the dynamics of a particle. A particle need not be either quantum or classical, but there is a third possible kind of dynamics, mesoscopic dynamics, which interpolates between quantum and classical. This dynamics is described by a nonlinear Schr\"{o}dinger equation [see Eqn. (\ref{nlse}) below]. The nonlinear term depends on the newly introduced parameter $\theta(m/m_{Pl})$, and its nature is such that the nonlinearity vanishes in the small mass limit, $m\ll m_{Pl}$, $\theta\rightarrow 1$. On the other hand the nonlinear Schr\"{o}dinger equation reduces to Newton's classical laws of motion in the limit $m\gg m_{Pl}$, $\theta\rightarrow 0$. This interpolating behaviour, where one makes a transition from quantum to classical mechanics via an intermediate nonlinear quantum mechanics, is not ruled out by experiment. Its verification or otherwise in the laboratory will constitute a crucial test of these ideas.

\section{The Cosmological Constant Problem}

The quantum-classical duality helps understand why there should be a cosmological constant of the order of the observed matter density; a possible explanation for the observed cosmic acceleration. 
If there is a non-zero cosmological constant term $\Lambda$ in the Einstein equations, of the standard form $\Lambda g_{ik}$, it follows from symmetry arguments that in the noncommutative generalization of gravity, a corresponding term of the form $\Lambda \theta_{ik}$ should also be present. This latter term vanishes in the macroscopic limit $m\gg m_{Pl}$ but is present in the microscopic limit $m\ll m_{Pl}$. 

However, when $m\ll m_{Pl}$, the effective gravitational constant goes to zero, so $\Lambda$ cannot be sourced by ordinary matter. Its only possible source is the zero-point energy associated with the
quantum particle $m\ll m_{Pl}$. Since this zero-point energy is necessarily non-zero, it follows that 
$\Lambda$ is necessarily non-zero.
This same $\Lambda$ manifests itself on cosmological scales,
where $\Lambda g_{ik}$ is non-vanishing, because $g_{ik}$ is non-vanishing, even though $\Lambda\theta$ goes to zero on cosmological scales, because $\theta$ goes to zero. Essentially we are saying that we have to examine the
two limits of $\Lambda(g_{ik}+\theta_{ik})$ : the microscopic limit and the macroscopic limit; the value of
$\Lambda$ arising at one of the limits will clearly be the same as its value at the other limit.

This solves the vexing problem of the cancellation of (i) a bare $\Lambda$ coming from general relativity, and
(ii) a $\Lambda$ coming from the zero point energy of quantum fields. This problem arises in the first place
because we have allowed ourselves to treat general relativity and quantum theory as completely disconnected
theories. The nonlinearity of the theory suggested here, the consequent duality, and the introduction of the antisymmetric component of the metric compel us to treat the two theories as limiting cases of an underlying theory, and to conclude that the so-called bare $\Lambda$ and the `quantum $\Lambda$' are
one and the same thing. The question of their mutual cancellation does not arise any longer. 

The value of $\Lambda$ can be estimated by appealing to the deduced quantum-classical duality.
The total mass in the observable Universe is $m_c\sim c^{3} (GH_0)^{-1}$, where $H_0$ is the present value of the Hubble constant. The mass dual to this $m_c$ is $m_q=m_{Pl}^2/m_c \sim hH_0/c^2$,
and $m_q c^2$ is roughly the magnitude of the zero-point energy in the ground state.
In a higher mode, the energy is a multiple of the ground state energy, and we write it as
$nm_{q}c^2\theta(nm_qc^2/m_{Pl})$, recalling that the effective Planck constant runs with energy.
To obtain a rough estimate, we take $\theta$ to be one for
energies up to $m_{Pl}$ and zero for energies beyond $m_{Pl}$. We then see that the total contribution
to the zero point energy is 
\begin{equation}
E_{tot}=hH_{0}[1+2+3+...+m_{Pl}c^{2}/hH_{0}]\sim  c^{5}H_{0}^{-1}/G.
\end{equation} 
It is remarkable that Planck's constant drops out of the sum! 
The vacuum energy density, and hence the value of the cosmological constant, is 
$c^{5}H_{0}^{-1}/ G (cH_{0}^{-1})^{3} \sim (cH_0)^2/G$ which is of the order of the observed value of $\Lambda$.

We note that the ground state energy $hH_{0}$ is
being mapped to a total energy $E_{tot}=(m_{Pl}c^{2})^{2}/hH_{0}$, which is an instance of a UV-IR mixing, or
equivalently, a quantum-classical duality.  As $H_{0}$ goes to zero, the IR limit goes to zero, whereas the
UV limit diverges.

Clearly, nothing in this argument singles out today's epoch; hence it follows that there
is an ever-present $\Lambda$, of the order $(cH)^2/G$, at any epoch, with $H$ being the Hubble constant at that epoch. This solves the cosmic coincidence and fine-tuning problems. However, issues related to an ever-present $\Lambda$ will have to be addressed - we will return to this aspect in the last section.

\bigskip

\noindent{\bf Understanding $\Lambda$} : The standard quantum field theoretic cosmological
constant problem does not arise here because we have brought in a new scale, the Hubble constant.
In effect, we are proposing that since $H_{0}^{-1}$ is the age of the Universe, there is a fundamental
minimum frequency, i.e. $H_{0}$. All allowed frequencies are discrete multiples of $H_{0}$, with the
maximum being at Planck frequency. As a result, the net zero point energy comes out to be $H_{0}^{-1}$.
By itself, this is higher than $m_{Pl}$; but we must recall that duality  demands this much to be the
classical contribution to the cosmological constant. Hence the energy density is found by dividing the       
total energy by the volume of the observed Universe, giving a value for $\Lambda$ that matches with observations.
Thus although the argument for obtaining the magnitude of $\Lambda$ given here draws input from quantum
theory, our argument is completely different in concept from what is suggested by quantum field theory.
For us, duality is playing a crucial role.

\section{Testing for dark energy through quantum measurement}

We would now like to suggest that the above proposal for the origin of the cosmological constant
can in principle be tested in the laboratory by examining the quantum mechanics of mesoscopic
systems, because the latter is also affected by the nonlinearity of the underlying theory.

Firstly, as discussed above, the effective Planck's constant is $\hbar\theta$, and using the form of $\theta$
that we have, we can write 
\begin{equation}
\hbar_{eff}=\hbar\theta=\frac{\hbar}{1+m/m_{Pl}}.
\end{equation}
Thus a measurement of the Planck constant for a `mesoscopic particle' with mass approaching
the Planck mass will show a deviation from the standard value. By particle we mean a composite
object in the required mass range whose internal degrees of freedom can be neglected.

Secondly, the nonlinearity can result in a breakdown of quantum superposition during a quantum
measurement, leading to collapse of the wave-function and a finite lifetime for superpositions.
A great deal has been written about the physics of quantum measurement over the last century or
so. It is fair to say that there are essentially only two possibilities : either the wave function
collapses during a quantum measurement, or it does not. If it does not, then the many worlds
interpretation holds, and the different worlds do not interfere because of decoherence. If the wave
function does collapse, then a modification of the Schr\"{o}dinger equation in the mesoscopic domain
is indicated. We have argued that the nonlinearity resulting from removal of external time favors the
collapse picture \cite{singh}.

As we discussed above, on the Planck mass/energy scale, the Klein-Gordon equation becomes nonlinear.
In the non-relativistic limit, it results in the following nonlinear Schr\"{o}dinger equation :
\begin{align} \begin{split}
i\hbar\frac{\partial\psi}{\partial t} &= -\frac{\hbar^{2}}{2m}\frac
{\partial^{2}\psi}{\partial x^{2}} + V(x)\psi\\\quad &+\frac{\hbar^{2}}{2m}(1-\theta)
\left(\frac{\partial^{2}\psi}{\partial x^{2}} - 
\left[\frac{\partial (\ln\psi)}{\partial x}\right]^{2}\psi
\right).
\label{nlse}
\end{split}\end{align}
This equation can be rewritten as
\begin{equation}
i\hbar\frac{\partial\psi}{\partial t} = -\frac{\hbar^{2}}{2m}
\frac{\partial^{2}\psi}{\partial x^{2}} + V(x)\psi + 
q\frac{\partial^{2}(\ln R)}{\partial x^{2}}\psi + 
i\frac{q}{\hbar}\frac{\partial^{2}\phi}{\partial x^{2}}\psi
\end{equation}
where $q=\frac{\hbar^{2}}{2m}(1-\theta)$ and $\psi=Re^{i\phi/\hbar}$.
Norm is preserved during evolution, provided the probability density is defined as 
$\rho=|\psi|^{2/\theta}$.

Since nonlinearity is negligible for the quantum system, prior to the onset of a quantum measurement, evolution is described by
\begin{equation}
i\hbar\frac{\partial\psi}{\partial t} = -\frac{\hbar^{2}}{2m}
\frac{\partial^{2}\psi}{\partial x^{2}} + V(x)\psi
\end{equation}
thus preserving superposition.
The onset of 
measurement corresponds to mapping the state $|\psi>$ to 
the state $|\psi>_F$ of the final system as
\begin{equation}
|\psi>\rightarrow |\psi>_F\ \equiv  \sum_n a_n|\psi>_{Fn}= \sum_n\ a_n|\phi_n>|A_n>
\label{map}
\end{equation}
where $|A_n>$ is the state the measuring apparatus would be in, had the 
initial  system been in the state $|\psi_n>$.

Evolution is now described by the equation
\begin{equation}
i\hbar\frac{\partial\psi_F}{\partial t} = H_F \psi_F +
q\frac{\partial^{2}(\ln R_F)}{\partial x^{2}}\psi + 
i\frac{q}{\hbar}\frac{\partial^{2}\phi_F}{\partial x^{2}}\psi_{F}
\end{equation}
where $q=\frac{\hbar^{2}}{2m_F}(1-\theta)$ and $\psi=Re^{i\phi_F/\hbar}$.
$m_{F}$ is the total mass of the final system, which includes the quantum system as
well as the measuring apparatus.
The states $\psi_{Fn}$ cannot evolve as a superposition because the
evolution is now non-linear.
However, the initial state at the onset of measurement \textit{is} a superposition of the $\psi_{Fn}$.
This superposition must thus break down during further evolution, according to the law
\begin{equation}
i\hbar\frac{\partial a_{n}}{\partial t} =  
i\frac{q_n}{\hbar}\frac{\partial^{2}\phi_F}{\partial x^{2}}a_{n}.
\label{supbreak}
\end{equation}
Note that the $q_n$'s have been set to be different for different states. This is to be expected because
$\theta$ will be determined by the quantum state, and setting it as a function only of $m/m_{Pl}$ to begin
with was a leading order approximation, applied for simplicity.
We thus get
\begin{equation}
\hbar^{2}\frac{d}{dt}\ln\frac{a_i}{a_j} = (q_i-q_j)\phi_F''
\end{equation}
and only the state with the largest $q$ survives \cite{singh}. In this manner, the inclusion of a nonlinear
term breaks superposition.

In order to recover the Born probability rule, it is essential that the $q_n$'s are random variables, with
a suitable probability distribution. Only further development in theory can determine if the $q_n$'s are indeed random, and if so, what their probability distribution is. A highly plausible candidate for a random variable 
is the phase of the quantum state at the onset of measurement. Although the phase evolves in a deterministic
manner, it is effectively random, because the time at which the measurement begins is arbitrary. 

From Eqn. (\ref{supbreak}) we can define the lifetime 
$\tau_{sup}$ of a superposition
\begin{equation}
\tau_{sup} = \frac{m}{(1-\theta)\phi_F''}. 
\label{sup1}
\end{equation}
Since $\theta$ is strictly equal to one in standard linear quantum mechanics,  quantum superposition has an infinite lifetime in the linear theory.
However, the situation begins to change in an interesting manner as the value of the mass
$m$ approaches and exceeds $m_{Pl}$. Since we know that in this limit $\theta$ approaches zero, we can
neglect $\theta$, and the superposition lifetime will then essentially be given by
\begin{equation}
\tau_{sup} \approx \frac{m}{\phi_F''} \sim \frac{mL^2}{\phi}. 
\label{sup2}
\end{equation}

We can get a numerical estimate by noting that we are close to the classical limit, where the phase coincides with
the classical action in the Hamilton-Jacobi equation. To leading order, the magnitude of the classical action is given by $S_{cl}=mc^{2}t$, where $t$ is the time over which we observe the classical trajectory; approximately, this could be taken to be the value of the phase $\phi$, and $\tau_{sup}$ is
then roughly given by
\begin{equation}
\tau_{sup}\sim \frac{1}{t}\left(\frac{L}{c}\right)^{2}.
\label{sup3}
\end{equation}
For a measuring apparatus, if we take the linear dimension to be say $1$ cm, and the time of observation to be say $10^{-3}$ seconds, we get the superposition lifetime to be $10^{-18}$ seconds.
We can get a very rough estimate of $\tau_{sup}$ for a mesoscopic system using
(\ref{sup3}), and taking $L\sim 10^{-3}$ cm, $m \sim 10^{-9}$ gm, $\phi\sim N\hbar$ with 
$N\sim 10^{15}$.
This gives $\tau_{sup}\sim 10^{-3}$ seconds. Thus an experimental detection of dependence of
superposition lifetime on the mass (equivalently number of degrees of freedom) of the system could
be indicative of the nonlinearity.

The third possible way in which a nonlinearity of this nature can be detected is through rapid
successive measurements of a quantum observable. Suppose a certain outcome $O_{1}$ for an observable
results from the random variable being in a certain range $\delta$. Suppose now that a second measurement is made 
sufficiently quickly with the eigenbasis slightly rotated. Because the random variable will not have changed
to a value sufficiently different from the original one, the result of the second measurement will show  
a correlation with the result of the first measurement, contrary to what standard quantum mechanics predicts.

A more detailed discussion of the physics of measurement described here
will be presented elsewhere \cite{ALS}.

\section{Can there be an ever present $\Lambda$?}

A positive cosmological `constant' which is of the order of $H^{2}$ at every epoch is obviously not
a constant, and does not satisfy the standard equation of state $p=-\rho$. Furthermore, by increasing
the rate of expansion in the early Universe, it spoils the consistency between theory and observation
with regard to the abundance of light elements. It also makes galaxy formation more difficult later during
the evolution of the Universe. It is thus evident that although the cosmic coincidence problem can be
solved by an ever-present $\Lambda$, one has to ascertain that the resulting cosmological model must be
consistent with observation. A way out, as has been suggested by Sorkin, is to have a cosmological constant 
whose mean value is zero, but which has fluctuations with a typical magnitude of the order of $H^{2}$
\cite{sorkin}. As a starting point, this seems like a reasonable possibility for us also, considering
that in our model the origin of $\Lambda$ lies in the zero-point energy contribution coming from quantum theory.
However, the development of a cosmological model in the context of our scenario is an issue we have not
yet addressed, and we leave this for future investigation.

A phenomenological model for an ever-present $\Lambda$ has been partially developed in the context of the causal
set approach to quantum gravity \cite{Ahmed}. In this approach, a fluctuating $\Lambda$ of the order $H^{2}$ 
is predicted because $\Lambda$ is conjugate to the spacetime four volume, and this volume itself is subject
to quantum fluctuations. The phenomenological model is specified by choosing a suitable equation of state for
$\Lambda$ and expressing $\Lambda$ as a stochastic function of the four volume. A numerical study by the authors 
shows tracking behavior in $\Lambda$, as well as fluctuations. For a suitable choice of a free parameter a 
$\Lambda$ consistent with the present observed value is reproduced. It has however been pointed out by 
Barrow \cite{Barrow} that the model is very strongly constrained by the magnitude of the CMB anisotropy
on the last scattering surface. It remains to be seen whether a way can be found out to overcome this constraint,
by constructing an inhomogeneous version of the phenomenological model, or otherwise. An alternative investigation
on the origin of $\Lambda$ based on quantum gravitational fluctuations has been carried out by
Padmanabhan \cite{Padmanabhan}.  

On a more general note we observe that the theoretical prediction of an ever-present non-zero cosmological `constant' of the order of $H^{2}$ is independent of the details of the cosmological model. Essentially, all we have assumed is a homogeneous and isotropic cosmology, but we have placed no a priori restrictions on the evolutionary history of the scale factor. Thus although we originally set out to seek an explanation for the observed cosmic acceleration in the framework of the standard Big Bang cosmology, we could turn things around and ask the following question : given an ever-present $\Lambda$, does it admit a non-standard cosmology consistent with observations? To us, the answer to this question is not obvious, and in our view the question merits further careful examination.  

\section{Concluding Remarks}
Our use of noncommutative spacetime has a conceptually different origin as compared to applications based
on the seminal work of Doplicher, Fredenhagen and Roberts \cite{DFR}. In the latter, spacetime noncommutation
relations are deduced as a consequence of the joint application of quantum uncertainty relations and the
rules of general relativity on the Planck length scale. One then envisages that quantum field theories
exhibit effects induced by these spacetime commutation relations, on the Planck length scale. Also, on these
scales general relativity could be assumed to be replaced by a noncommutative gravity theory which should
eventually be quantized.

For us, the starting point has been that there should be a reformulation of quantum mechanics which does
not refer to a classical time. This leads to the conclusion that linear quantum theory is a limiting  case
of an underlying theory which becomes nonlinear on the Planck energy scale. This is the principle difference
from the theories referred to in the previous paragraph - the latter assume a strict validity of linear quantum
theory at all scales. For us, this nonlinearity is responsible for the explanation of the tiny observed
cosmological constant, and possibly also the collapse of the wave-function during a quantum measurement.
In order to arrive at the proposed reformulation of quantum mechanics we are led to suggest noncommutativity
not only in spacetime, but also in momentum space. While the detailed theory remains to be developed, some
consequences of the heuristic discussions given here can be tested in the laboratory.
 
\bigskip

\noindent{\bf Acknowledgments :} I would like to thank Aruna Kesavan,
Kinjalk Lochan and Aseem Paranjape for useful discussions.

\end{document}